# Suppression of reflection from the grid boundary in solving the time-dependent Schrödinger equation by split-step technique with fast Fourier transform

A. A. Gonoskov[*] and I. A. Gonoskov[**]

*Institute of Applied Physics, Russian Academy of Sciences, Nizhny Novgorod 603950, Russia*

We present an approach to numerically solving the time-dependent Schrödinger equation and other parabolic equations by the split-step technique with fast Fourier transform, which suppresses the backreflection of waves from the grid boundaries with any specified accuracy. Most importantly, all known methods work well only for a narrow region of incident waves spectrum, and the proposed method provides absorption of any wave whose length is large enough in comparison with the size of absorption region.

[*] Electronic address: argon1@mail.ru
[**] Electronic address: gonoskov@ufp.appl.sci-nnov.ru

For a number of reasons, the split-step technique with fast Fourier transform (FFT) is often most efficient for numerically solving the parabolic equations and the time-dependent Schrödinger equation, as a special case. Here the multi-dimensional grid appears to be closed. In a 2D case, for example, we have a toroidal topology. This means that if some wave packet comes to the grid boundary it will pass through and appear at the opposite side. However, when solving a physical problem, one often deals with a part of infinite space. This raises the problem of absorbing the backreflection of all waves from the grid boundaries. Traditionally, this problem is solved by using an imaginary potential in some absorption region at the grid boundary (see, for example, [2]). Parameters of the imaginary potential are chosen experimentally, depending on which waves should be absorbed. Recent paper [3] has reported some modification of this method, which, however, does not make it free from a significant drawback. It appears that this method provides relatively good results only for a narrow region of incident waves spectrum. Short waves pass through the absorption region, whereas long waves are reflected. Here the size of the absorption region should considerably exceed the absorbed wavelength. In addition, parameters of the imaginary potential need to be re-chosen for each new problem. In this paper, we propose a technique that provides absorption of any wave whose length is large enough in comparison with the size of the absorption region. In the multidimensional case it means that one can use this method to absorb waves with different angles of incidence. The analogous problem for electromagnetic waves has been already solved with perfectly matched layer (PML) technique [5]. Most importantly, in the proposed method it is possible to derive a precise formula for calculating errors and to determine the applicability domain.

For simplicity, we will consider a one-dimensional case. Let us assume a one-dimensional grid, on which we solve the time-dependent dimensionless Schrödinger equation:

$$i\frac{\partial \psi(x,t)}{\partial t} = -\frac{1}{2}\frac{\partial^2}{\partial x^2}\psi(x,t) + U(x,t)\psi(x,t). \quad (1)$$

The calculations are made using the split-step technique [1]:

$$\psi(x, t+\Delta t) = e^{-iU(x,t+\Delta t)\frac{\Delta t}{2}} \times \hat{F}^{-1}\left(e^{-ik^2\frac{\Delta t}{2}}\hat{F}\left(e^{-iU(x,t)\frac{\Delta t}{2}}\psi(x,t)\right)\right) + O(\Delta t^3), \quad (2)$$

where $\hat{F}$ is the Fourier transform operator. All waves traveling to the boundary of the grid should be absorbed and, besides, the wave function should be affected only in a small region. Let us consider a region of length $L_0$ with the center at point $X_0$, at which incoming broad-spectrum wave packet has to be absorbed.

If we use in this grid an imaginary potential in the form, e.g., $U_{ap}(x) = i \cdot m \cdot \cos^2\left(\pi \frac{x-X_0}{L_0}\right)$, then we obtain an absorption of incoming wave packet [2]. It appears, however, that by choosing parameter $m$, we can have satisfactory absorption only for a narrow band of the wave packet spectrum, corresponding to waves with a wavelength exceeding $L_0$. Short waves pass through the absorption region, whereas long waves are reflected. This gives rise to an idea of a method in which the absorption coefficient $m$ is dependent on the wavelength. Let us expand the wave function in the region $\left[X_0 - \frac{L_0}{2}; X_0 + \frac{L_0}{2}\right]$ in a Fourier integral:

$$\psi(x,t) = \int_{-L_0/2\pi}^{L_0/2\pi} C_k(t) e^{ik(x-X_0)} dk. \quad (3)$$

Since equation (1) is linear, the absorption should also be linear. In a general case, it may be described by the following formula:

$$\frac{1}{C_k}\frac{\partial C_k}{\partial t} = S(|k|), \quad (4)$$

where $S(|k|)$ is a function that determines the suppression method. However, we will consider a specific case with the following considerations. The propagation velocity of waves with a wave number $k$ is proportional to $k$. The wave packet which has the mean velocity $k$ will pass through the absorption region in time proportional to $k^{-1}$ and according to (4) its amplitude will exponentially decrease while it passes through the region. If we want to affect equally all wave packets we should choose the rate of amplitude decreasing proportional to $k$. Therefore, the absorption rate of $C_k$ should be proportional to $k$. Thus we obtain the following form of the equation describing the absorption:

$$\frac{1}{C_k}\frac{\partial C_k}{\partial t} = -\beta|k|, \quad (5)$$

where $\beta$ is an absorption rate coefficient. In this approach, however, we will have a sharp jump because of the disagreement between wave function values at points at the absorption region boundaries, resulting in the formation of a broad spectrum of wave packet, although in the whole space the wave packet traveling to the absorption region may have a narrow spectrum. Thus, the spectrum will contain harmonics of the waves non-existent in the whole space. It is clear, that the suppression of a wave may be interpreted as a generation of the same wave with the opposite phase. Thus, absorption of non-existent waves will efficiently lead to waves generation and, consequently, to weak absorption, as we observed in numerical experiment. To avoid this, it is reasonable to affect only a part of the wave function by using a "mask" evenly smoothing the edges to zero:

$$\psi(x,t) = \psi(x,t) \cdot (1 - f(x)) + \psi(x,t) f(x), \quad (6)$$

where function $f(x)$ (mask) should smoothly decay from unity to zero at the boundaries of the absorption region. The suppression technique should be used for Fourier expansion coefficients of the function $\Phi(x,t) = \psi(x,t) f(x)$. For best results, the function $f(x)$ should have as narrow spectrum as possible. This condition is fulfilled for function $\cos^2\left(\pi \frac{x - X_0}{L_0}\right)$. Now it is obvious that if we use equation $\frac{1}{C_k}\frac{\partial C_k}{\partial t} = -m$ instead of equation (5), we will obtain the imaginary potential method. The only parameter $\beta$ in equation (5) is responsible for the wave damping factor. The integration of (5) gives:

$$\beta \sim \frac{1}{L_0} \ln\left(\frac{|\psi_i|}{|\psi_f|}\right), \quad (7)$$

where $\psi_i$ and $\psi_f$ are amplitudes of wave packet before and after passing through the absorption region. Thus, the parameter $\beta$ should be chosen based on the required accuracy of calculations. It was found in numerical experiments (Fig. 1) that the absorption result is well described by formula (7) for absorbed wavelengths not exceeding a certain value that decreases with the parameter $\beta$.

Further we will describe in brief the algorithm scheme. During each iteration, in addition to the step operation of the split-step method, the following procedures should be performed:

1. Present the wave function $\psi(x,t)$ as a sum of two parts by applying the mask in the absorption region:
    $\psi(x,t) = \Psi(x,t) + \Phi(x,t)$,
    where:
    $\Psi(x,t) = \psi(x,t)(1 - f(x))$,
    $\Phi(x,t) = \psi(x,t) f(x)$.

2. Find Fourier coefficients using FFT for $\Phi(x,t)$ in the absorption region:
    $$\Phi(x,t) = \sum_{n=-N/2+1}^{n=N/2} C_n(t) e^{in\frac{2\pi}{L_0}(x - X_0)}.$$

3. Perform one iteration for coefficients $C_n$:
    $$C_n(t + \Delta t) = C_n(t) \cdot e^{-\beta \frac{2\pi}{L_0}|n|\Delta t}.$$

4. Perform the inverse FFT:
    $$\overline{\Phi(x,t)} = \sum_{n=-N/2+1}^{n=N/2} C_n(t + \Delta t) e^{in\frac{2\pi}{L_0}(x - X_0)}.$$

5. Obtain the wave function in the absorption region after absorption:
    $\overline{\psi(x,t)} = \Psi(x,t) + \overline{\Phi(x,t)}$.

Note that due to the FFT the speed of this procedure, as well as the whole split-step method with FFT, is proportional to $V \log_2(V)$, where $V$ is the number of matrix elements.

Since the Schrödinger equation and the absorption methods are linear, for comparison it is enough to determine the coefficients of reflection and transmission of a test wave through the absorption region for different wavelengths. So, we need to generate the test wave with different values of the wavelength.

Below we discuss how to solve this problem. The problem of wave generation is equal to the absorption problem. The wave function should be affected in some region ("generation" region) by using the absorption method for difference between the wave function and the analytically described test wave. For this purpose, during each iteration, the following procedures should be performed:

1. In the wave generation region, determine the difference between the wave function and the analytically described test wave:
    $$G(x,t) = \psi(x,t) - A e^{ikx - i\frac{k^2}{2}t},$$
    where $A$ and $k$ are parameters of the test wave that corresponds to the dispersion relation for equation (1) in free space.

2. Apply the absorption method in the generation region for $G(x,t)$:
    $$G(x,t) \xrightarrow{absorption} \overline{G(x,t)}.$$

3. Obtain the wave function in the generation region after generation:
    $$\overline{\psi(x,t)} = \overline{G(x,t)} + A e^{ikx - i\frac{k^2}{2}t}.$$

It is clear, that in the generation region these procedures make the wave function close to the test wave. As the test wave corresponds to the dispersion relation for equation (1) in free space, the test wave goes out of the generation region. So we get the wave generation. These procedures also provide absorption of all other waves which come to the generation region.

Below we will present the results of comparison of this method with the imaginary potential method. To compare the methods, the test wave was generated using the imaginary potential method. Some time after the start of generation, a picture was set in, where the transmitted and reflected waves could be distinguished. Below we plot (Fig. 1) the dependences of the sum of squares of reflection and transmission coefficients of the test wave (T and D) on the log of the ratio of the absorption region length $L_0$ to the wavelength $\lambda$ for the both methods (the imaginary potential method with three different values of the parameter $m$ and the proposed method with coefficients $\beta = 0.48$ and $\beta = 0.69$).

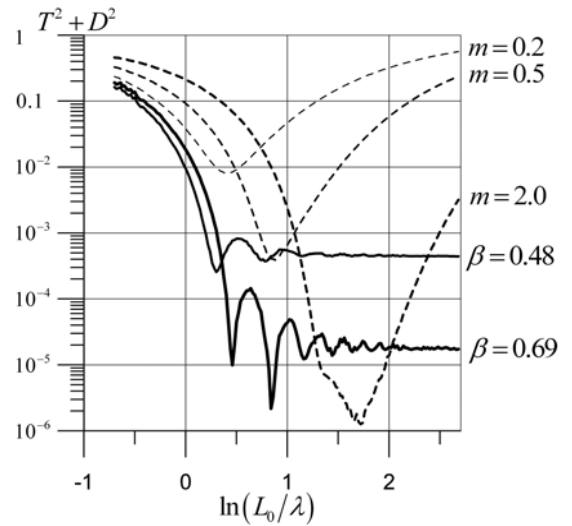

FIG. 1. The dependences of the sum of squares of reflection and transmission coefficients of the test wave (T and D) on the log of the ratio of the absorption region length $L_0$ to the test wavelength $\lambda$. Imaginary potential method – dotted lines; proposed method – full lines.

It can be seen that the parameter $\beta$ in the proposed method can be chosen to provide better result, as compared to the imaginary potential method, for all wavelengths. Note that the imaginary potential method works very poorly for high and low frequencies with the exception of the narrow region centered at the definite value of $L_0/\lambda$, depending on amplitude $m$, whereas the proposed method is free of this feature and in the high-frequency region it provides stable absorption of waves, corresponding to formula (7). This stable absorption begins already for wavelengths that only 1.3 – 1.5 times exceed the length of the absorption region (for $\beta = 0.48$).

The comparison with the approach proposed in [3] (referred below as "filtered") is shown in Fig. 2. Here $a$ is a filter absorption coefficient, $b$ is an imaginary potential coefficient. It can be also seen that the proposed method, by appropriate choice of the parameter $\beta$, can be made better than the filtered method for all wavelengths.

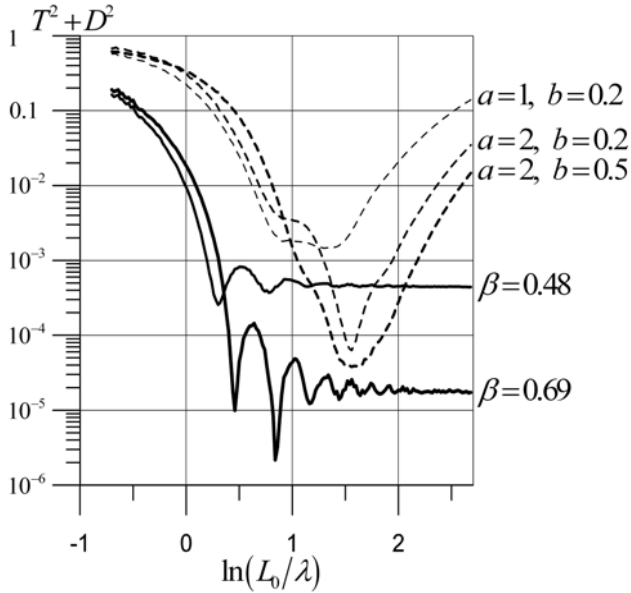

FIG. 2. The dependences of the sum of squares of reflection and transmission coefficients of the test wave (T and D) on the log of the ratio of the absorption region length $L_0$ to the test wavelength $\lambda$. Filtered method – dotted lines; proposed method – full lines.

Let us now discuss additional possibilities provided by the proposed method for solving the scattering problem. The method can be easily generalized for a multi-dimensional case. For this we should use the one-dimensional procedure independently for every line, column, etc., for each dimension. However, when numerically solving the scattering problem, one can significantly improve the results by further modifying the proposed technique.

The scattering problem is put as follows. For simplicity, we will consider a 2D case. We assume that there is a 2D grid with a certain distribution of potential that comes to a constant value at the grid boundary. In the example given below, the potential has the form $U(\vec{r}) = \dfrac{1}{|\vec{r}|}$. An incident wave, corresponding to the dispersion relation of the Schrödinger equation

$$i\frac{\partial \psi(\vec{r},t)}{\partial t} = -\frac{1}{2}\nabla^2 \psi(\vec{r},t) + U(\vec{r})\psi(\vec{r},t) \quad (8)$$

in free space, hits the potential from the left. When numerically solving the Schrödinger equation (8), it is required to calculate in the whole space the wave function of the electron scattered by the potential.

The numerical experiment can be arranged as follows. Initially, the wave function of the electron is zero on the grid. At the grid boundary, this function is affected by using the above-described method for the difference between the wave function of the electron and the analytically described plane wave, which we want to make the incident wave. As a result, we will get a generation of the incident wave at the left boundary. It is clear that first the incident wave front will enter the grid and then it will be changed when passing the potential. Some time later, the fronts of transmitted and reflected waves will reach the grid boundary and the probability distribution will become static. The time dependence will persist only as a phase change.

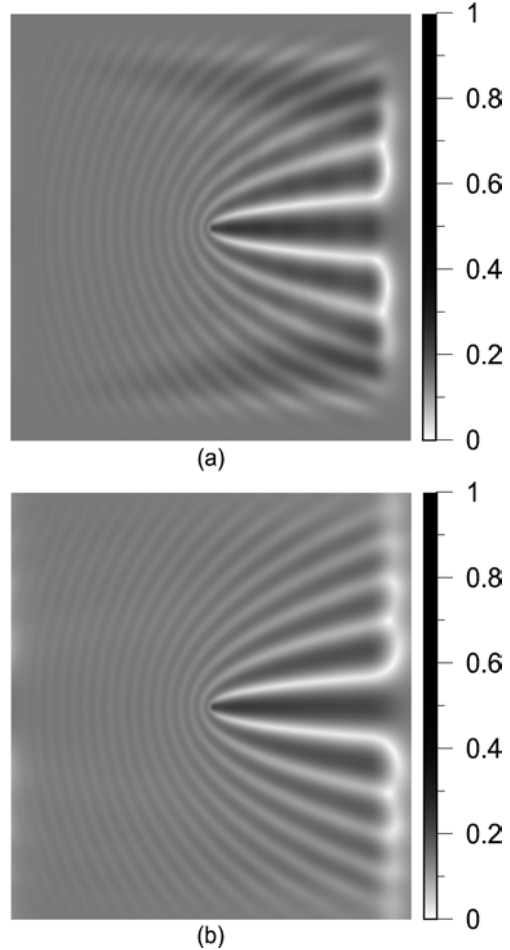

FIG. 3. Probability distribution in 2D scattering problem. (a) – the imaginary potential method; (b) – the proposed method.

The result of the numerical experiment conducted using the imaginary potential technique is shown in Fig. 3(a). One can see dark regions at the upper and lower boundaries, which get wider to the right boundary. These perturbations occurring above and below correspond to reflections from the upper and lower boundaries. This is due to the following fact. Far away from the center of the potential, the analytic solution for the scattering problem is the sum of two parts, a scattered wave and the incident plane wave. If the potential decays insufficiently fast, the incident plane wave has an additive to the phase, which is proportional to log of $|\vec{r}|$ (for Coulomb potential, see [4]) even far away from the center of the potential. As a result, at the upper and lower boundaries we suppress a wave that differs from a required one only by phase. This corresponds to generation of one more wave that differs only by phase. As a result of the interference of these waves, we obtain the distribution shown in Fig. 3(a). To avoid this, it is possible, for example, not to absorb a wave traveling along the upper and lower boundaries, i.e., to absorb only a scattered wave. This means that in the proposed method waves that are long in the transverse direction should not be absorbed. To achieve this, one should use expression:

$$\frac{1}{C_k}\frac{\partial C_k}{\partial t} = -\beta\begin{cases}|k|, |k| > 1 \\ 0, |k| \leq 1\end{cases} \quad (9)$$

rather than the expression (5), thereby performing absorption with wavelength discrimination. Thus there will be no absorption for coefficients $C_k$ with $|k|$ lower or equal to unity which correspond to waves that are the longest in the transverse direction. The distribution provided by this modification is shown in Fig. 3(b). We can see that there are no dark regions, which were in Fig. 3(a). The figure shows that the area of the grid where the wave function is free of parasitical distortions is about two times larger than in the case of imaginary potential method Fig. 3(a). That means that the use of this modification of the proposed method instead of the imaginary potential method for the 2D scattering problem allows to get the same results on two times smaller grid. Moreover, in the 3D scattering problem this modification allows to use four times smaller grid.

It should be noted in conclusion that suggested method can be used for solving any parabolic equations by the split-step technique with fast Fourier transform. It is important that after appropriated choosing of the parameters it works better than well-known imaginary potential method for any wave packets which should be absorbed. Also this method allows to selectively affect the different harmonics of the wave packet. This feature can be very useful for some physical problems. Besides, the proposed method is very good for waves generation. In general, the proposed technique allows not only to significantly reduce the size of the absorption region, but to research some new problems when numerically solving the parabolic equations.

Authors would like to thank M. Yu. Ryabikin for useful comments. We acknowledge a support from the Dynasty Foundation and RFBR (grant N05-02-17523).